\documentclass[aps,pra,twocolumn,floatfix]{revtex4-2}
\usepackage{graphics}% Include figure files
\usepackage{bm}% bold matha
\usepackage[caption=false,labelformat=simple]{subfig}% For subfigure
\usepackage[dvips]{graphicx}
\usepackage{amsmath}
\usepackage{amssymb}
\usepackage{blkarray}
\usepackage[dvipsnames]{xcolor}
\usepackage[colorlinks=true,linkcolor=blue]{hyperref}
\usepackage[normalem]{ulem}

\renewcommand{\d}{{d}}
\begin{document}
\draft

\title{Magnetic Feshbach resonances between atoms in $^2$S and $^3$P$_0$ states: \hfil\break
mechanisms and dependence on atomic properties}

\author{Bijit Mukherjee}
\affiliation{Joint Quantum Centre (JQC) Durham-Newcastle, Department of
Chemistry, Durham University, South Road, Durham, DH1 3LE, United Kingdom.}
\author{Matthew D. Frye}
\affiliation{Joint Quantum Centre (JQC) Durham-Newcastle, Department of
Chemistry, Durham University, South Road, Durham, DH1 3LE, United Kingdom.}
\author{Jeremy M. Hutson}
\email{j.m.hutson@durham.ac.uk} \affiliation{Joint Quantum Centre (JQC)
Durham-Newcastle, Department of Chemistry, Durham University, South Road,
Durham, DH1 3LE, United Kingdom.}

\date{\today}

\begin{abstract}
Magnetically tunable Feshbach resonances exist in ultracold collisions between atoms in $^2$S and $^3$P$_0$ states, such as an alkali-metal atom colliding with Yb or Sr in a clock state. We investigate the mechanisms of these resonances and identify the terms in the collision Hamiltonian responsible for them. They involve indirect coupling between the open and closed channels, via intermediate channels involving atoms in $^3$P$_1$ states. The resonance widths are generally proportional to the square of the magnetic field and are strongly enhanced when the magnitude of the background scattering length is large. For any given pair of atoms, the scattering length can be tuned discretely by choosing different isotopes of the $^3$P$_0$ atom. For each combination of an alkali-metal atom and either Yb or Sr, we consider the prospects of finding an isotopic combination that has both a large background scattering length and resonances at high but experimentally accessible field. We conclude that $^{87}$Rb+Yb, Cs+Yb and $^{85}$Rb+Sr are particularly promising combinations.
\end{abstract}

\maketitle

\section{Introduction}

There is currently growing interest in forming ultracold molecules containing an alkali-metal atom and an atom such as Sr, Yb or Hg with a closed-shell ground state. These molecules are paramagnetic in nature and have strong electric dipole moments. Such molecules have potential applications in quantum many-body systems \cite{Micheli:2006}, quantum simulation and quantum information \cite{Perez_Rios:2010, Herrera:2014}, precision measurement \cite{Alyabyshev:2012}, testing of fundamental symmetries \cite{Meyer:2009} and tuning of collisions and chemical reactions \cite{Abrahamsson:2007, Quemener:2016}.

There are two approaches that have been successful in producing samples of ultracold molecules. A few molecules, such as CaF and SrF, have been cooled to ultracold temperatures by a combination of helium buffer-gas cooling and laser cooling \cite{Tarbutt:laser-cooling:2018}. However, this approach is applicable only for molecules with near-diagonal Franck-Condon factors, which permit a nearly closed cycle of laser absorption and emission. A wider range of ultracold molecules have been produced from ultracold atoms, by either photoassociation or magnetoassociation, followed by laser transfer to the rovibrational ground state. In magnetoassociation, molecules are produced by sweeping a magnetic field adiabatically across a magnetic Feshbach resonance, where a molecular state crosses a state of the free atomic pair as a function of magnetic field \cite{Hutson:IRPC:2006, Chin:RMP:2010}. This approach has now been applied to many different molecules, though it is limited to those formed from atoms that are themselves coolable to ultracold temperatures.

Mixtures of alkali-metal and closed-shell ($^1$S) atoms possess magnetic Feshbach resonances, but they are sparse in magnetic field, often occurring at magnetic fields that are too high to be experimentally accessible. They are also very narrow, because the molecular state is coupled to the atomic pair state by only the distance-dependence of the atomic hyperfine interaction \cite{Zuchowski:RbSr:2010, Brue:LiYb:2012, Brue:AlkYb:2013, Yang:CsYb:2019}. Much experimental work has been devoted to locating and observing Feshbach resonances of this type \cite{Nemitz:2009, Baumer:2011, Muenchow:2011, Borkowski:2013, Ivanov:2011, Hansen:2011, Roy:2016, Guttridge:2017, Guttridge:1p:2018, Guttridge:2p:2018, Barbe:RbSr:2018, Green:2019, Green:LiYb-res:2020, Franzen:CsYb-res:2022}. They have now been observed for both bosonic and fermionic isotopes of Sr interacting with Rb \cite{Barbe:RbSr:2018}, for fermionic $^{173}$Yb interacting with $^6$Li \cite{Green:LiYb-res:2020} and for $^{173}$Yb interacting with Cs \cite{Franzen:CsYb-res:2022}. However, attempts to use them to form ultracold molecules by magnetoassociation have so far been unsuccessful.

We have recently proposed using alkaline-earth atoms \footnote{Atoms such as Yb and Hg are grouped along with alkaline-earth atoms because of their closed-shell ground states and similar pattern of excited states to Ca and Sr.} in their excited $^3$P states, rather than their ground states, for initial molecule formation \cite{Mukherjee:RbYb:2022}. Ultracold samples of Sr and Yb are readily prepared in their metastable $^3$P$_2$ and $^3$P$_0$ states. An atom in a $^3$P state interacts with one in a $^2$S state to form multiple molecular electronic states, and the resulting spin-dependent and anisotropic interactions provide additional mechanisms for Feshbach resonances. For Rb+Yb($^3$P), we carried out coupled-channel calculations as a function of magnetic field, to characterize the magnetic Feshbach resonances and the bound and quasibound states associated with them. For Rb+Yb($^3$P$_2$), we showed that broad Feshbach resonances exist, but that molecules formed at them are likely to decay in a few microseconds to form Yb atoms in $^3$P$_0$ and $^3$P$_1$ states. For Rb+Yb($^3$P$_0$), however, the molecules are stable except for decay to Yb($^1$S) combined with Rb($^2$S) or Rb($^2$P), and such processes are expected to be much slower. The resonances at the Rb+Yb($^3$P$_0$) threshold are typically broader than those for Rb+Yb($^1$S), and there are additional resonances involving rotationally excited (d-wave) molecular states. At the $^3$P$_0$ threshold, by contrast with $^1$S, resonances due to d-wave states can exist even for isotopes of Sr or Yb without nuclear spin.

The purpose of the present paper is to investigate the mechanisms responsible for resonances between atoms in $^2$S and $^3$P$_0$ states, and explore how the resonances depend on the properties of the atoms. The hyperfine splitting of the $^2$S atom and the spin-orbit splitting of the $^3$P atom vary greatly between species. As a result, the levels that can cause Feshbach resonances have different vibrational quantum numbers, and the couplings between the states have quite different effects.

The structure of the paper is as follows. Section \ref{sec:theory} introduces the interaction potentials, describes our theoretical methods, and considers the selection rules for different terms in the collision Hamiltonian. Section \ref{sec:widths} investigates the dependence of the widths on magnetic field, atomic properties, and background scattering lengths, for resonances due to both s-wave and d-wave bound states. Section \ref{sec:likely} considers which combinations of alkali-metal and $^3$P$_0$ atoms are most likely to have resonances wide enough to observe at experimentally accessible fields. Finally, Section \ref{sec:conclude} presents conclusions and perspectives.

\section{Theory} \label{sec:theory}

\subsection{Electronic structure and potential curves}

We consider the interaction of an atom in a $^2$S state (atom 1) with one in a $^3$P state (atom 2). In the absence of spin-orbit coupling, the interaction produces 4 spin-orbit-free electronic states of symmetry $^2\Sigma$, $^2\Pi$, $^4\Sigma$ and $^4\Pi$, labeled by their spin multiplicity $2S+1$ and the projection $\Lambda=0\ (\Sigma)$ or $\pm1\ (\Pi$) of the orbital angular momentum onto the molecular axis. The spin-orbit interaction splits the $^3$P state of the isolated atom into 3 components with total electronic angular momentum $j=0$, 1 and 2. At short range, it splits the molecular $^2\Pi$ state into components with $\Omega=1/2$ and 3/2 and the $^4\Pi$ state into components with $\Omega=1/2$, 1/2, 3/2 and 5/2, with 1/2 appearing twice because $\Omega$ is the \emph{magnitude} of the projection of the total electronic angular momentum (orbital and spin) onto the molecular axis. There are a total of 9 electronic states arising from $^2$S+$^3$P: 5 with $\Omega=1/2$, 3 with $\Omega=3/2$ and one with $\Omega=5/2$. The spin-orbit coupling mixes different states with different $S$ and/or $\Lambda$ but the same $\Omega$.

\begin{figure}[tbp]
	\subfloat[]{
		\includegraphics[width=0.45\textwidth]{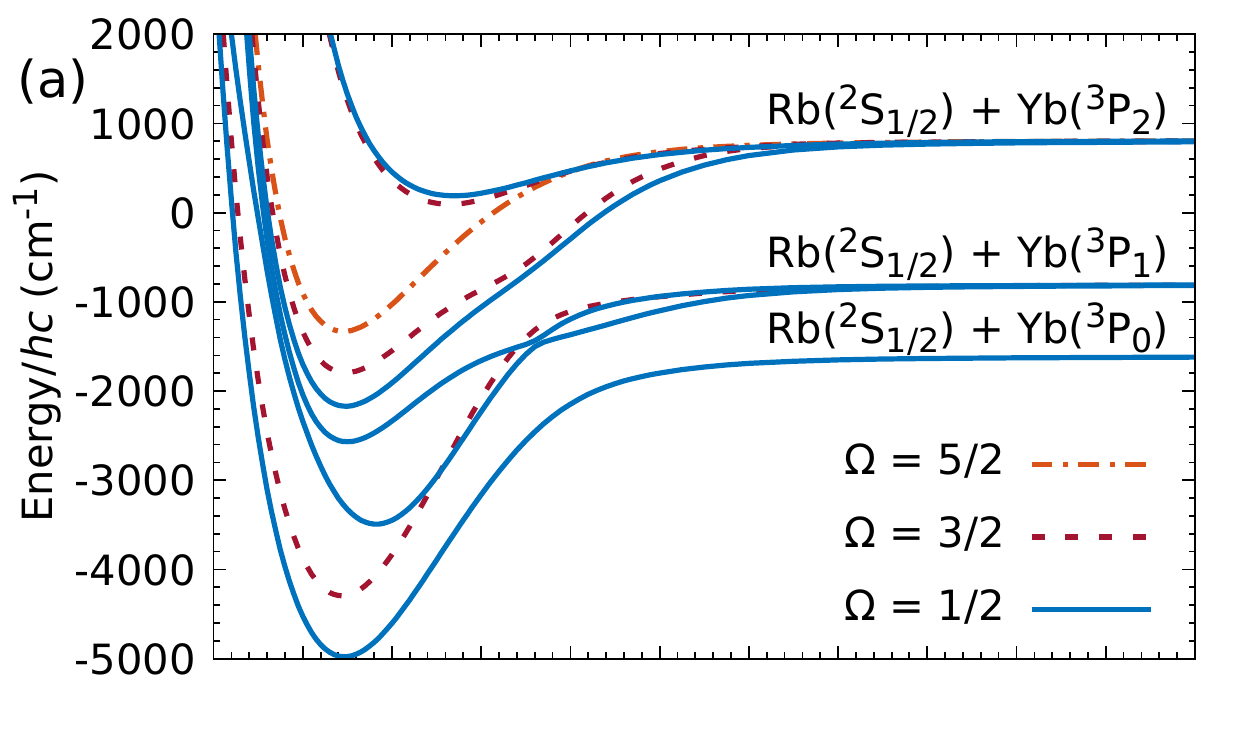}
	}
    \vspace{-0.8 cm}
	\subfloat[]{
		\includegraphics[width=0.45\textwidth]{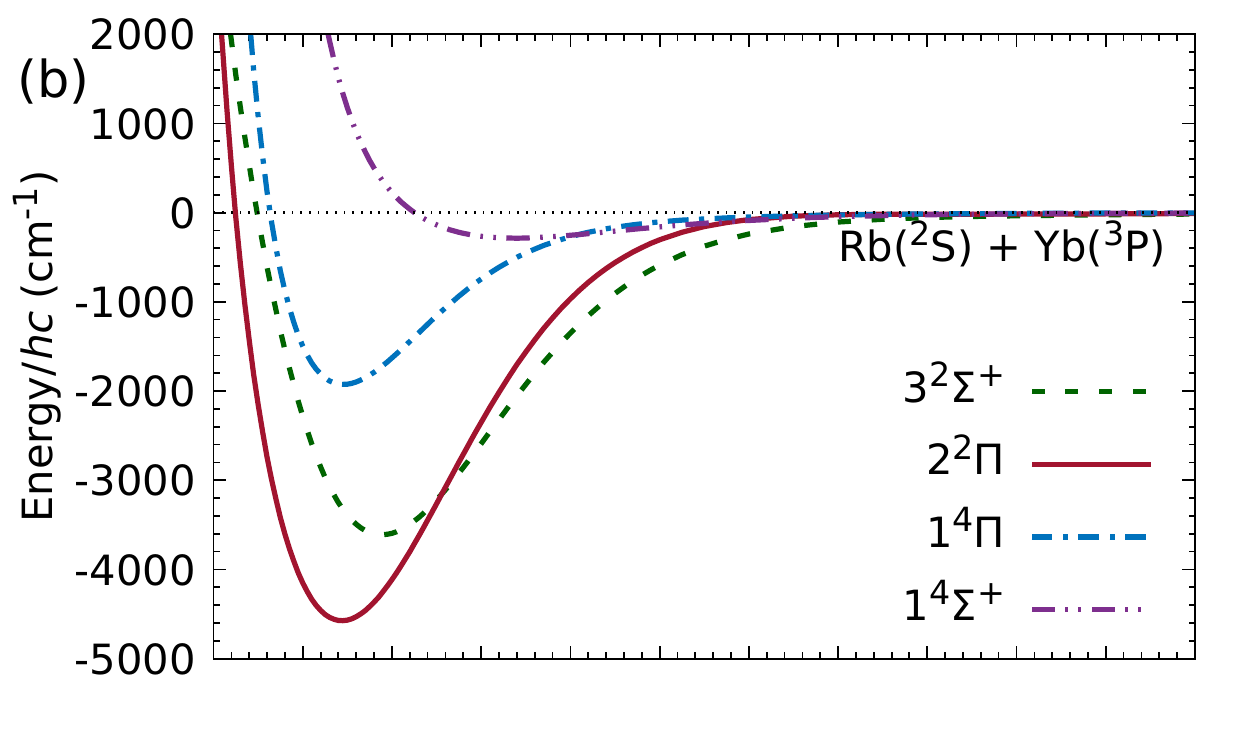}
	}
	\vspace{-0.8 cm}
	\subfloat[]{
		\includegraphics[width=0.45\textwidth]{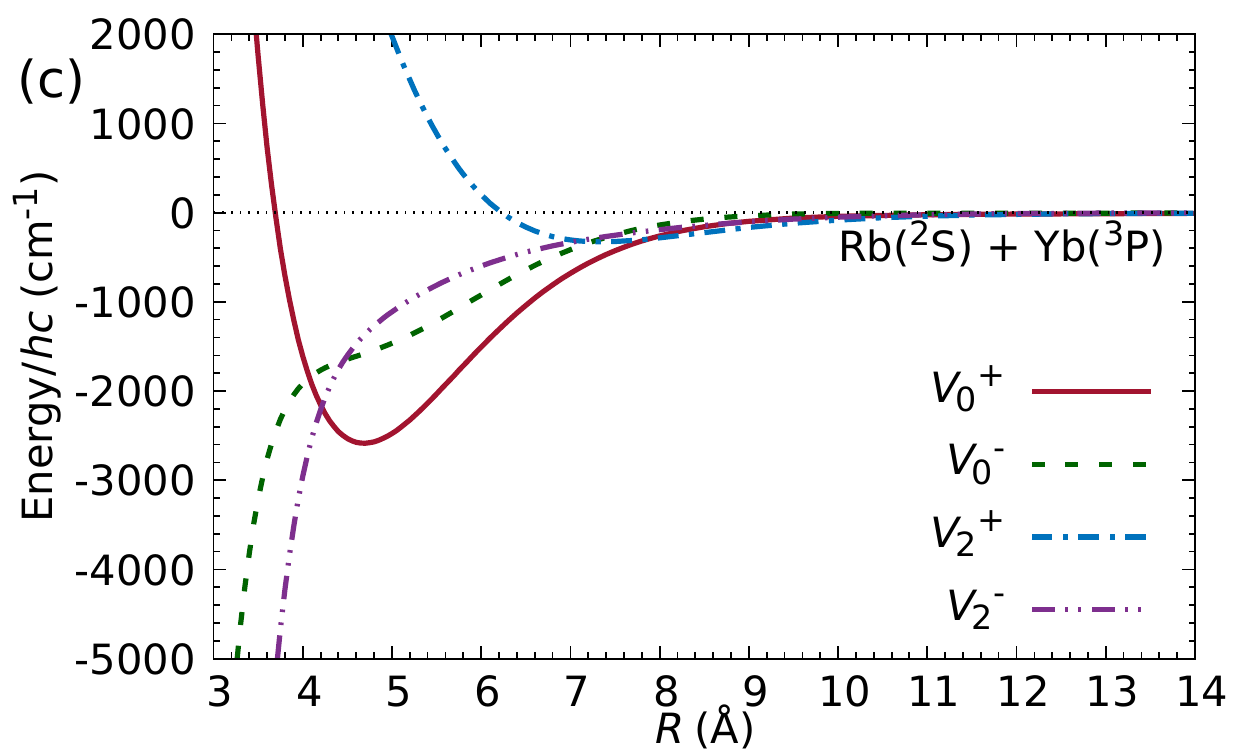}
	}
    \vspace{-0.4 cm}
    \caption{Potential-energy curves for Rb+Yb($^3$P) \cite{Mukherjee:RbYb:2022}. (a) Adiabatic curves including spin-orbit coupling;
    (b) spin-orbit-free curves;
    (c) spin-averaged and spin-difference curves.
    The $R$-dependent spin-orbit coupling functions used in Ref.\ \cite{Mukherjee:RbYb:2022} are not used in the present work, so are not shown.
}%
\label{fig:pot-curves}
\end{figure}

Coupled-channel calculations on these systems require couplings between the electronic states as well as the potential curves themselves. For this, it is most convenient to define potential curves for the $^2\Sigma$, $^2\Pi$, $^4\Sigma$ and $^4\Pi$ states, excluding spin-orbit coupling so that $S$ and $\Lambda$ are conserved. The couplings between the states are then provided by separate spin-orbit coupling operators, which may depend on the internuclear distance $R$.
Suitable spin-orbit-free curves have been obtained directly from correlated electronic-structure calculations for the ground and excited states of Li+Yb \cite{Zhang:2010} and Rb+Sr \cite{Zuchowski:2014, Pototschnig:RbSr:2014}. For Rb+Yb \cite{Shundalau:2017} and Cs+Yb \cite{Meniailava:2017}, the published curves include spin-orbit coupling, and in Ref.\  \cite{Mukherjee:RbYb:2022} we fitted the curves for Rb+Yb to obtain spin-orbit-free curves and $R$-dependent spin-orbit couplings; the results are shown in Fig.\ \ref{fig:pot-curves} for illustration. The spin-orbit-free curves for the other systems considered here show the same general features.

The electronic interaction operator may be written
\begin{equation}
\hat{V}_\textrm{elec}(R,\xi) = \sum_{S,\Lambda} V^S_\Lambda(R) \hat{\cal V}^S_\Lambda(\xi) + \hat{V}_\textrm{so}(R,\xi),
\label{eq:elec-int}
\end{equation}
where $\xi$ represents the electronic orbital and spin degrees of freedom. $\hat{V}_\textrm{so}(R,\xi)$ represents an $R$-dependent contribution to the spin-orbit coupling, excluding the contribution from the free atom. The operators $\hat{\cal V}^S_\Lambda(\xi) = | ^{2S+1}\Lambda \rangle \langle ^{2S+1}\Lambda |$ project onto states with well-defined values of $S$ and $\Lambda$.

To understand the couplings responsible for Feshbach resonances, it is useful to represent the interaction potentials for $\Sigma$ and $\Pi$ states in terms of isotropic and anisotropic components, $V_0$ and $V_2$, for each total spin, $S=1/2$ (doublet) and $S=3/2$ (quartet) \cite{DUBERNET:open:1994},
\begin{align}
\label{eq:v02}
V_0^S(R) &= \frac{1}{3} \left( V_\Sigma^S(R) + 2V_\Pi^S(R) \right) \\
V_2^S(R) &= \frac{5}{3} \left( V_\Sigma^S(R) -  V_\Pi^S(R) \right).
\end{align}
It is also useful to define averages and differences of the potential curves for the doublet and quartet states, $S=1/2$ and 3/2,
\begin{equation}
V^\pm_\kappa(R) = \frac{1}{2} \left( V_\kappa^{1/2}(R) \pm V_\kappa^{3/2}(R) \right),
\end{equation}
where $\kappa = 0,2$. The resulting combinations of potential curves for Rb+Yb are shown in Fig.\ \ref{fig:pot-curves}(c).
The corresponding operators may be written $\hat{V}^\pm_\kappa(R,\xi)$, with spin-dependent parts $\hat{\cal V}^\pm_\kappa(\xi)$, so that the interaction operator of Eq.\ (\ref{eq:elec-int}) is
\begin{align}
\hat{V}_\textrm{elec}(R,\xi) & = \sum_{\pm,\kappa=0,2} \hat{V}^\pm_\kappa(R,\xi) + \hat{V}_\textrm{so}(R,\xi) \nonumber\\
& = \sum_{\pm,\kappa=0,2} V^\pm_\kappa(R) \hat{\cal V}^\pm_\kappa(\xi) + \hat{V}_\textrm{so}(R,\xi).
\label{eq:vpm}
\end{align}
This representation is convenient because the operators $\hat{\cal V}^\pm_\kappa(\xi)$ have simple selection rules in the angular-momentum basis sets described below.

\subsection{Coupled-channel methods} \label{sec:cc}

The Hamiltonian of the interacting pair of atoms is
\begin{align}
\hat H &=-\frac{\hbar^2}{2\mu}R^{-1}\frac{\d^2\ }{\d R^2}R
+ \frac{\hbar^2 \hat L^2}{2\mu R^2} + \hat{H}_1 + \hat{H}_2 \nonumber\\
&+ \hat{V}_\textrm{elec}(R,\xi) +\hat{V}^\textrm{d}(R,\xi).
\label{eqh}
\end{align}
Here the first term describes the kinetic energy with respect to the internuclear distance $R$, while $\hbar^2 \hat L^2/2\mu R^2$ is the centrifugal term that describes the end-over-end rotation of the interacting pair. $\hat{H}_1$ and $\hat{H}_2$ are the Hamiltonians of the isolated $^2$S and $^3$P atoms, including Zeeman terms for a magnetic field $B$ along the $z$ axis,
\begin{align}
\hat{H}_1 &= \zeta_1 \hat{i}_1\cdot\hat{s}_1 + \left( g_{s,1} \hat{s}_{1,z} + g_i\hat{i}_{1,z} \right) \mu_\textrm{B} B; \label{eq:h2S} \\
\hat{H}_2 &= a_2 \hat{l}_2\cdot\hat{s}_2 + a_2^{(1)} \delta_{j1}
+ \left( \hat{l}_{2,z} + g_{s,2}\hat{s}_{2,z} \right) \mu_\textrm{B} B.
\label{eq:h3P}
\end{align}
Here $\hat{s}_1$ and $\hat{i}_1$  are the vector operators for the electron and nuclear spin angular momenta of the $^2$S atom, with components along $z$ indicated by subscript $z$, and $\zeta_1$ is its hyperfine coupling constant.
$\hat{l}_2$ and $\hat{s}_2$ are the vector operators for the electron orbital and spin angular momentum of the $^3$P atom, $a_2$ is its spin-orbit coupling constant, and $a_2^{(1)}$ is a correction (usually small) for the effects of $jj$-coupling.
$g_{s,1}$ and $g_{s,2}$ are the g-factors for the electron spins, and $g_i$ is that for the nuclear spin of the $^2$S atom.
$\hat{V}_\textrm{elec}(R,\xi)$ is the electronic interaction operator of Eq.\ \ref{eq:vpm}, and $\hat{V}^\textrm{d}(R,\xi)$ represents the magnetic dipole-dipole interaction between the electron spins on the two atoms. A non-zero nuclear spin for the $^3$P atom and $R$-dependent contributions to hyperfine couplings could be added straightforwardly, but are not included in the present work.

We carry out coupled-channel bound-state and scattering calculations as described in Ref.\ \cite{Mukherjee:RbYb:2022}. The bound-state calculations are performed using the packages \textsc{bound} and \textsc{field} \cite{bound+field:2019, mbf-github:2020}, which converge upon bound-state energies at fixed field, or bound-state fields at fixed energy, respectively. The methods used for bound states are described in Ref.\ \cite{Hutson:CPC:1994}. The scattering calculations are performed using the package \textsc{molscat} \cite{molscat:2019, mbf-github:2020}, which has special features for automatically converging on and finding the parameters of Feshbach resonances. All three packages use the same plug-in routines for basis sets and interaction operators. The basis sets, propagators and convergence parameters used here are the same as in Ref.\ \cite{Mukherjee:RbYb:2022}.

\subsection{Angular momenta and selection rules} \label{sec:basis}

There are 5 sources of angular momentum in a colliding pair, with quantum numbers $s_1$, $i_1$, $l_2$, $s_2$ and $L$, corresponding to the operators defined above. The separated atoms are best represented by quantum numbers $(s_1,i_1)f,m_f$ and $(l_2,s_2)j,m_j$, where the notation $(a,b)c$ indicates that $c$ is the resultant of $a$ and $b$ and $m_c$ is the projection of $c$ onto the $z$ axis.

To understand the couplings that produce Feshbach resonances, we use a basis set of eigenfunctions of the atomic Hamiltonians, Eqs.\ \ref{eq:h2S} and \ref{eq:h3P}, and $\hat{L}^2$. These are field-dressed functions, but they are conveniently expanded in zero-field functions $|f,m_f\rangle$, $|j,m_j\rangle$ and $|L,M_L\rangle$. Since the Zeeman terms are diagonal in $L$, $M_L$, $m_f$ and $m_j$ and have matrix elements off-diagonal in $j$ or $f$ by only $\pm1$, the expansions are simple.

All the operators in the Hamiltonian (\ref{eqh}) conserve $M_\textrm{tot}=m_f+m_j+M_L$.
$\hat{V}_0^+$ is entirely diagonal.
$\hat{V}_0^-$ can change $f$ and/or $j$ by 1; it can change $m_f$ and $m_j$ by $0,\pm1$ while conserving $m_f+m_j$.
$\hat{V}_2^+$ can change $j$ and/or $L$ by 0 or 2 but has no matrix elements from 0 to 0; it can change $m_j$ and $M_L$ by $0,\pm1,\pm2$ while conserving $m_j+M_L$.
$\hat{V}_2^-$ combines the selection rules of $\hat{V}_0^-$ and $\hat{V}_2^+$, conserving only $M_\textrm{tot}$.

\section{Dependence of resonance widths on atomic properties and magnetic field}
\label{sec:widths}

When a molecular bound state crosses a scattering threshold as a function of magnetic field, it produces a magnetically tunable Feshbach resonance. In the absence of inelastic scattering, this is characterized by a pole in the s-wave scattering length $a(B)$,
\begin{equation}
a(B)=a_\textrm{bg}\left(1-\frac{\Delta}{B-B_\textrm{res}}\right),
\label{eq:aFR}
\end{equation}
where $B_\textrm{res}$ is the position of the resonance, $\Delta$ is its width and $a_\textrm{bg}$ is a background scattering length that varies slowly with $B$.

Fermi's Golden Rule gives an approximate expression for the energy width $\Gamma_E$ of a state that lies above an open channel,
\begin{equation}
\Gamma_E=2\pi\left|\langle\alpha,n|\hat H'(R,\xi)|\beta,k\rangle\right|^2.
\label{eq:goldenrule}
\end{equation}
Here $\langle \alpha,n|$ represents a bound state $n$ in closed channel $\alpha$, $|\beta,k\rangle$ represents a scattering state with wavevector $k$ in open channel $\beta$ and $\hat{H}'(R,\xi)=H'(R)\hat{\cal H}'(\xi)$ is the operator that couples the channels and causes the resonance.
The scattering state is normalized to a $\delta$-function of energy and has asymptotic amplitude $(2\mu/\pi\hbar^2 k)^{1/2}$. At limitingly low collision energy $E_\textrm{coll}=\hbar^2 k^2/2\mu$, $\Gamma_E$ depends on $k$ as \cite{Mies:Feshbach:2000}
\begin{equation}
\Gamma_E(k) \xrightarrow{k\to 0} 2ka_\textrm{bg}\Gamma_0,
\label{eq:gamma0}
\end{equation}
where $\Gamma_0$ is independent of energy. The width of the resonance in $a(B)$ is
\begin{equation}
\Delta=\frac{\Gamma_0}{\delta\mu_{\rm res}},
\label{eq:GammaToDelta}
\end{equation}
where $\delta\mu_{\rm res}$ is the difference between the magnetic moment of the molecular bound state and that of the free atom pair.
The expression for $\Delta$ factorizes into spin-dependent and radial terms,
\begin{equation}
\Delta = \frac{\pi \left|\langle\alpha|\hat{\cal H}'|\beta\rangle\right|^2 I_{nk}^2}{k a_\textrm{bg}\delta\mu_\textrm{res}},
\label{eq:FGRDelta1}
\end{equation}
where
\begin{equation}
I_{nk} = \int_0^\infty\psi_{n}(R) H'(R) \psi_{k}(R)\,{\rm d}R
\label{eq:Ink}
\end{equation}
and $R^{-1}\psi_n(R)$ and $R^{-1}\psi_k(R)$ are the radial parts of the bound-state and scattering wavefunctions.
It should be noted that the Golden Rule may produce overestimates of the widths in cases where $\hat{H}'(R,\xi)$ is too large to act perturbatively.

The factor of $a_\textrm{bg}$ in the denominator of Eq.\ \ref{eq:FGRDelta1} produces large values of $\Delta$ when $a_\textrm{bg}$ is small. However, these large values are unphysical, because the strength of the pole in Eq.\ \ref{eq:aFR} is actually $a_\textrm{bg}\Delta$. It is thus convenient to define a normalized width $\bar{\Delta}=(a_\textrm{bg}/\bar{a})\Delta$ \cite{Yang:CsYb:2019}, where $\bar{a}$ is the mean scattering length of Gribakin and Flambaum \cite{Gribakin:1993}, which is 82.8~$a_0$ for $^{87}$Rb+$^{174}$Yb.

If the effective interaction potentials for the incoming and resonant channels are closely parallel to one another, and the operator $\hat{H}'(R,\xi)$ acts principally at short range, $I_{nk}$ produces a simple dependence of $\Delta$ on $a_\textrm{bg}$ for the incoming channel and the binding energy $E_n$ of the bound state with respect to the threshold that supports it \cite{Brue:AlkYb:2013}. The dependence on $a_\textrm{bg}$ can be explained with quantum defect theory (QDT). Near threshold, the amplitude of the scattering wavefunction is proportional to the QDT function $C(k)^{-1}$ \cite{Mies:1984a}. The width is thus proportional to $C(k)^{-2}$, which near threshold is \cite{Mies:MQDT:2000, Julienne:2006}
\begin{equation}
C(k)^{-2} = k\bar{a}\left[1+\left(1-\frac{a_\textrm{bg}}{\bar{a}}\right)^2\right].
\label{eq:cqdt}
\end{equation}
The function has a minimum value of $k\bar{a}$ when $a_\textrm{bg}=\bar{a}$ but is approximately $ka_{\rm bg}^2/\bar{a}$ when $|a_\textrm{bg}|\gg\bar{a}$.
The dependence on $E_n$ arises from the behavior of near-dissociation vibrational states \cite{LeRoy:1970}. For an interaction potential that varies asymptotically as $-C_6 R^{-6}$, as here, the amplitude of the wavefunction at short range is proportional to $E_n^{1/3}$, so that $\Delta$ is proportional to $E_n^{2/3}$ \cite{Brue:AlkYb:2013}.

\subsection{Resonant states with $L_\textrm{res}=0$}

\begin{figure}[tbp]
\centering
\includegraphics[width=0.5\textwidth]{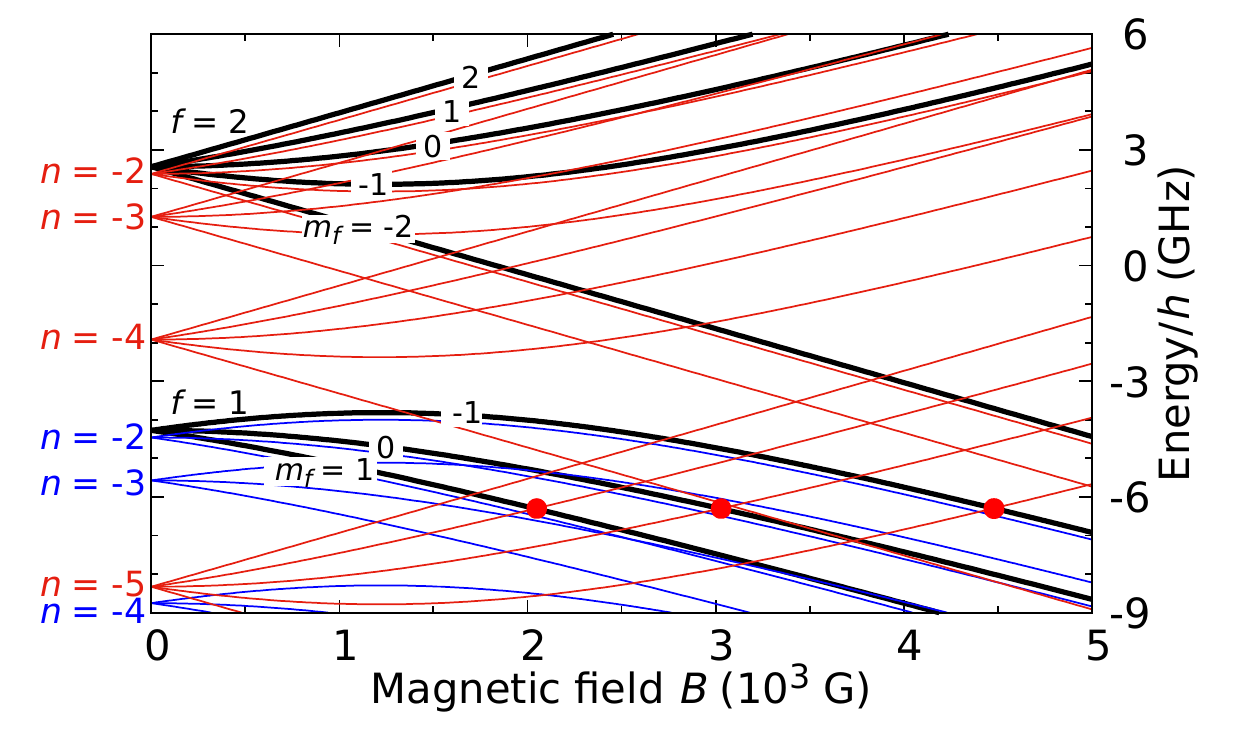}%
\caption{Level diagram for $^{87}$Rb + $^{174}$Yb($^3$P$_0$) with the unscaled interaction potential. The heavy black lines show the hyperfine thresholds for $f = 1$ and 2. Molecular levels with $L=0$ are shown in blue for $f=1$ and red for $f=2$, labeled with vibrational quantum number $n$. Crossings that cause Feshbach resonances are indicated with circles. Levels with $n=-1$ are too close to threshold to be visible. }%
\label{fig:crossing-L0}
\end{figure}

In considering resonance widths, it is helpful to identify quantum numbers associated with the incoming channel by a subscript `in' and those associated with the resonant bound or quasibound state by a subscript `res'. In the present work we are concerned with s-wave scattering, $L_\textrm{in}=0$.

A typical level diagram showing thresholds and potentially resonant states for $^{87}$Rb+Yb($^3$P$_0$) is shown in Fig.\ \ref{fig:crossing-L0}. The energies of the thresholds for $f=1$ and 2 are shown as black lines. Each threshold supports a set of near-threshold vibrational levels, labeled by quantum number $n=-1$, $-2$, counted downwards from the threshold that supports them. The rotationless states (s-wave states, $L_\textrm{res}=0$) supported by the thresholds for $f=1$ are shown as blue lines. There is only very weak mixing between different channels with $j=0$, and each bound or quasibound state is very nearly parallel to the threshold that supports it as a function of magnetic field $B$. Resonant states with quantum numbers ($f_\textrm{res},m_{f,\textrm{res}}$) thus do not cross thresholds with ($f_\textrm{in},m_{f,\textrm{in}}) = (f_\textrm{res},m_{f,\textrm{res}}$).
Such states can cross thresholds with $f_\textrm{in} = f_\textrm{res}$ but $m_{f,\textrm{in}} \ne m_{f,\textrm{res}}$. However, when $j=0$ and $L=0$, $M_\textrm{tot}=m_f$; no combination of the operators above can couple channels with different $M_\textrm{tot}$, so such crossings do not produce Feshbach resonances.

There are also states arising from the upper hyperfine threshold, $f=2$ for $^{87}$Rb, shown as red lines in Fig.~\ref{fig:crossing-L0}. For the unscaled potential shown in Fig.\ \ref{fig:pot-curves}, the states with $f=2$ and $n=-5$ lie about 11~GHz below the thresholds for $f=2$ at zero field, so 4~GHz below the thresholds for $f=1$. States with ($f_\textrm{res},m_{f,\textrm{res}}$) can cross thresholds with $f_\textrm{in} \ne f_\textrm{res}$ but $m_{f,\textrm{in}} = m_{f,\textrm{res}}$, and these crossings do produce Feshbach resonances (red circles in Fig.\ \ref{fig:crossing-L0}).

\begin{figure}[tbp]
\centering
\includegraphics[width=0.5\textwidth]{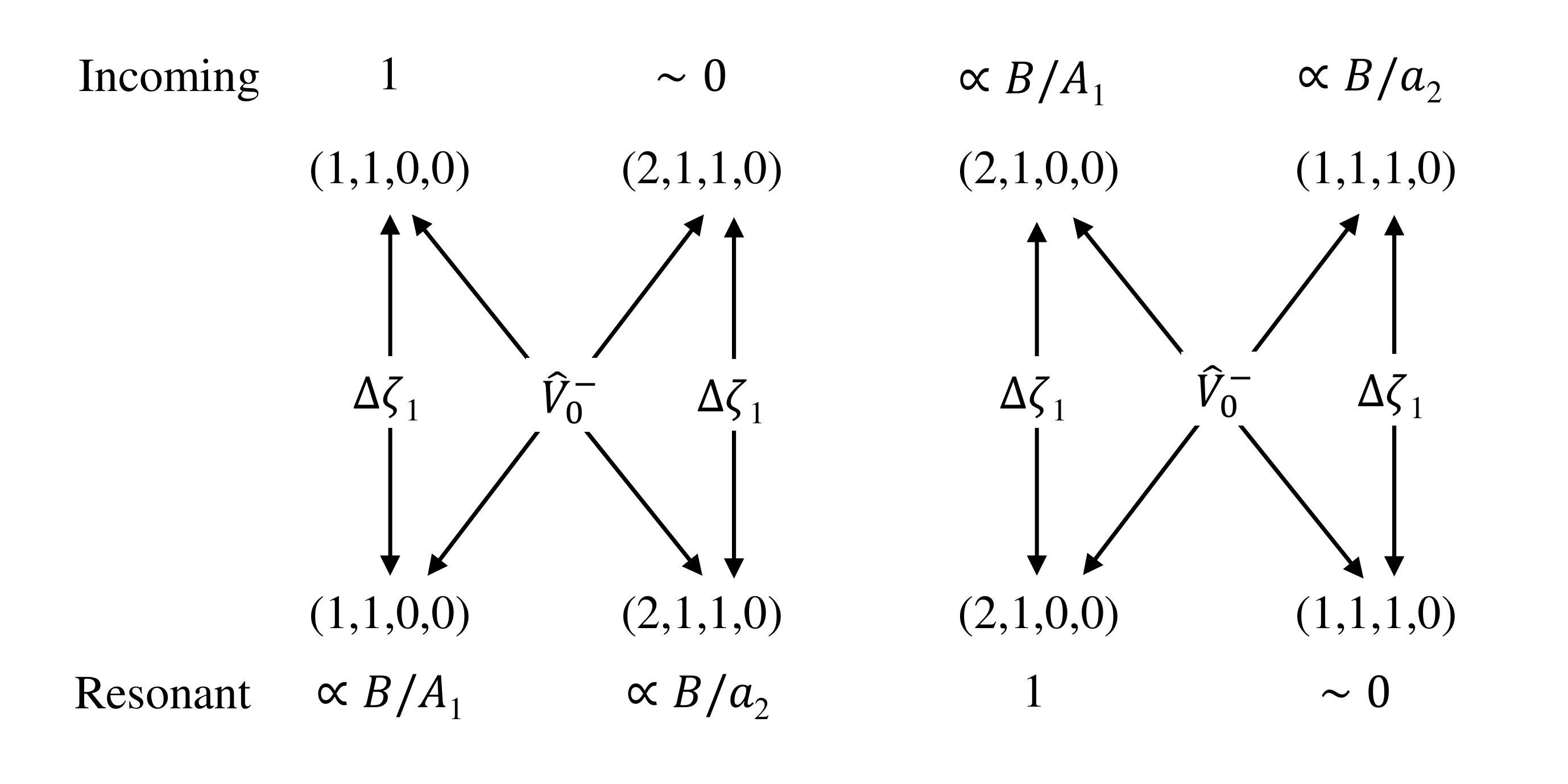}
\caption{Mechanism of the resonance in the incoming channel $(f_\textrm{in},m_{f,\textrm{in}})=(1,1)$ for $^{87}$Rb+Yb($^3$P$_0$), due to a bound state with $(f_\textrm{res},m_{f,\textrm{res}})=(2,1)$ and $L_\textrm{res}=0$. Components of the channel functions, designated by $(f,m_f,j,m_j)$, are shown for the incoming and resonant channels, together with the matrix elements between them that drive the resonance.}
\label{fig:coupling-L0}
\end{figure}

To understand the coupling responsible for these resonances, consider the channel basis functions involved at the incoming and resonant thresholds. For example, for $^{87}$Rb+Yb at the lowest threshold, the Rb function is predominantly $(f,m_f)=(1,1)$, but with an admixture of (2,1) proportional to $B/A_1$ at low field, where $A_1=\zeta_1(i_1+\frac{1}{2})$ is the hyperfine splitting. The Yb function is predominantly $(j,m_j)=(0,0)$, with an admixture of (1,0) proportional to $B/a_2$. The channel function for the pair thus has 4 components $(f,m_f,j,m_j)$, as shown in the top row of Fig.\ \ref{fig:coupling-L0}; the component proportional to $B^2$ is small at low fields. Conversely, the channel function for the resonant channel is made up of the same 4 components, but with different amplitudes as shown in the bottom row of Fig.\ \ref{fig:coupling-L0}. The operator $\hat{V}_0^-$ has direct matrix elements between the (1,1,0,0) and (2,1,1,0) components and between the (1,1,1,0) and (2,1,0,0) components, as shown by the diagonal arrows in Fig.\ \ref{fig:coupling-L0}. These provide an overall matrix element between the incoming and resonant channels proportional to $B/a_2$ in the Golden Rule expression (\ref{eq:goldenrule}).
There is no $B$-independent contribution to the matrix element because $f$, $j$ and $L$ couple to produce the total angular momentum ${\cal F}$, which is conserved at zero field; for $L=0$, ${\cal F}_\textrm{in}=1$ and ${\cal F}_\textrm{res}=2$, so no $B$-independent term can couple them.
The resulting widths $\Gamma_E$ and $\Delta$ are proportional to $B^2$ at low field and approximately inversely proportional to $a_2^2$, although the latter breaks down for small $a_2$ because $\hat{V}_0^-$ is too strong to act perturbatively at short range. None of the other operators $\hat{V}_2^+$, $\hat{V}_2^-$ or $\hat{V}^\textrm{d}$ have any matrix elements between the functions of Fig.\ \ref{fig:coupling-L0}. The operator $\hat{V}_\textrm{so}$ can also have matrix elements between the same components as $\hat{V}_0^-$, but they are usually much smaller.

There are additional couplings between the functions of Fig.\ \ref{fig:coupling-L0} due to the $R$-dependence of the hyperfine coupling. This is characterized by an additional interaction operator of the form $\Delta\zeta_1(R) \hat{i}_1\cdot\hat{s}_1$, where
\begin{equation}
\zeta_1(R) = \zeta_1 + \Delta\zeta_1(R).
\label{eq:zetaR}
\end{equation}
This is exactly analogous to Mechanism I for interaction of an atom in a $^2$S state with one in a $^1$S state \cite{Zuchowski:RbSr:2010, Barbe:RbSr:2018, Yang:CsYb:2019} which is due to a similar term $\Delta\zeta_1(R)$ in the hyperfine coupling. These matrix elements are included as vertical arrows in Fig.\ \ref{fig:coupling-L0}, but for $^3$P$_0$ states $\hat{V}_0^-$ is much stronger.

\begin{figure}[tbp]
\centering
\includegraphics[width=0.5\textwidth]{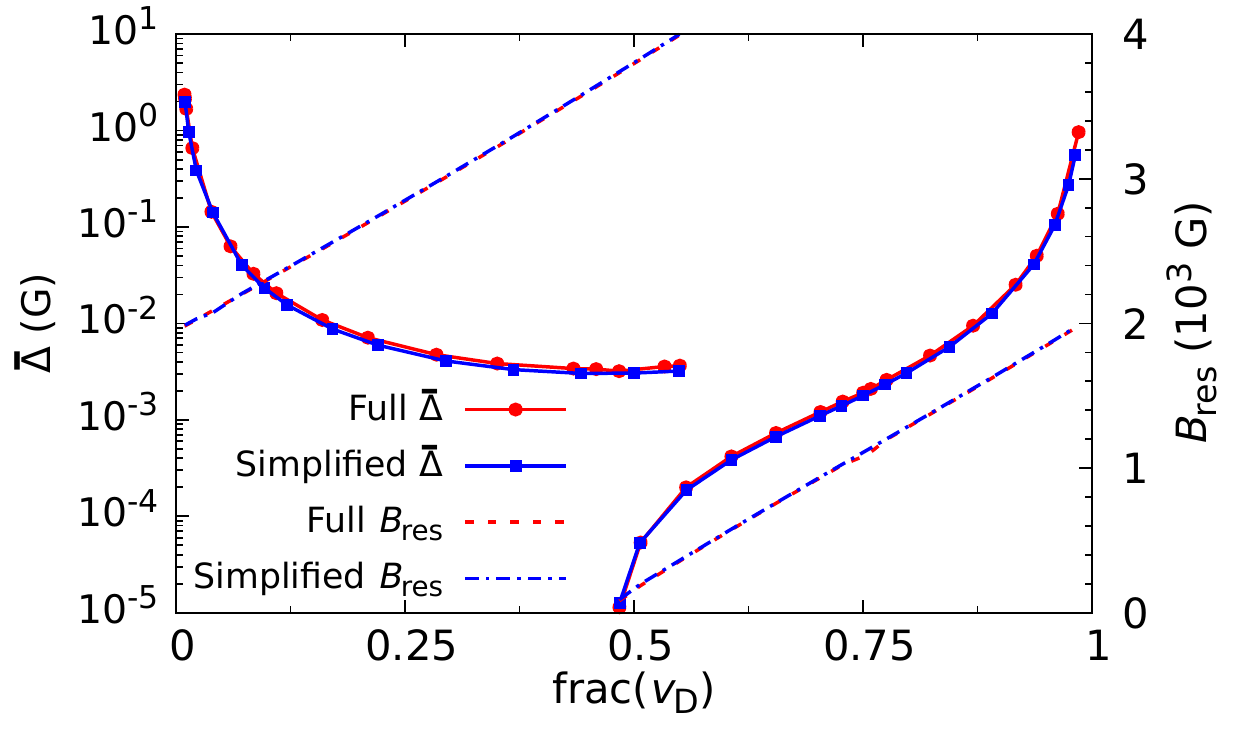}
\caption{Positions $B_\textrm{res}$ (dashed lines) and normalized widths $\bar{\Delta}$ (symbols) for resonances with $L_\textrm{res}=0$, using the full (red) and simplified (blue) Hamiltonians. The quantity frac($v_\textrm{D}$) is related to $a_\textrm{bg}$ through Eq.\ \ref{eq:vD}. The positions and widths are calculated for $^{87}$Rb+$^{174}$Yb, but would be very similar for a different isotope of Yb for an interaction potential scaled to give the same $a_\textrm{bg}$. }
\label{fig:simplified-L0}
\end{figure}

To verify that couplings involving $\hat{V}_0^-$ are dominant, we have carried out coupled-channel calculations using a simplified Hamiltonian that omits $\hat{V}_2^+$, $\hat{V}_2^-$, $\hat{V}_\textrm{so}$ and $\hat{V}^\textrm{d}$, leaving only $\hat{V}_0^+$, $\hat{V}_0^-$, and the $R$-independent atomic terms (\ref{eq:h2S}) and (\ref{eq:h3P}). Figure \ref{fig:simplified-L0} compares the position and width of the resonance at the lowest threshold, obtained from coupled-channel calculations using the full and simplified Hamiltonians; they are shown as a function of the background scattering length in the incoming channel, which in each case is adjusted with a small overall scaling $\lambda_\textrm{scl}$ of the interaction potential (less than 1\% change) as in Ref.\ \onlinecite{Mukherjee:RbYb:2022}.
Since $a_\textrm{bg}$ passes through a pole as a function of $\lambda_\textrm{scl}$, we linearize the horizontal axis by plotting the curves as a function of the fractional part of the vibrational quantum number at threshold \cite{Lutz:non-BO:2016},
\begin{equation}
\textrm{frac}(v_\textrm{D})=\frac{\tan^{-1}(1-a_\textrm{bg}/\bar{a})}{\pi}+\frac{1}{2};
\label{eq:vD}
 \end{equation}
this is $0$ when $a_\textrm{bg}=\infty$ and $\frac{1}{2}$ when $a_\textrm{bg}=\bar{a}$. The resonance positions and widths are very similar for the full and simplified Hamiltonians.

The resonance position depends on $a_\textrm{bg}$ because the zero-field binding energies $E_n$ are functions of $a_\textrm{bg}$. If $\lambda_\textrm{scl}$ is chosen so that $a_\textrm{bg}=\infty$, and small shifts due to $\hat{V}_0^-$ are neglected, the least-bound state ($n=-1$) for each channel lies exactly at the threshold for that channel, with binding energy $E_{-1}=0$. If $\lambda_\textrm{scl}$ increases, so that the potential becomes deeper, $a_\textrm{bg}$ decreases from $+\infty$ and all the states $n$ become more deeply bound. As $\lambda_\textrm{scl}$ increases further, $a_\textrm{bg}$ approaches $-\infty$ and passes through a pole; a new state then enters the potential from above, and the cycle repeats. For $^{87}$Rb+Yb($^3$P$_0$), the zero-field states with $f_\textrm{res}=2$, $n=-5$ cross the thresholds for $f_\textrm{in}=1$ when $a_\textrm{bg}=92.5\ a_0$. At that point the states with $m_{f,\textrm{res}}=+1$, 0 and $-1$ all cross the corresponding thresholds with $f_\textrm{in}=1$ at $B=0$. As $a_\textrm{bg}$ decreases from this value, the binding energies increase, so that the crossings shift to higher field.

The resonance width depends on $a_\textrm{bg}$ through a combination of several effects. First, it is approximately proportional to $B_\textrm{res}^2$ because of the contribution of channels with $j=1$ as described above. Secondly, it depends on $a_\textrm{bg}$ through Eq.\ \ref{eq:cqdt}, with a pronounced peak around $a_\textrm{bg}=\infty$, corresponding to frac($v_\textrm{D})=0$. Lastly, there is a (relatively weak) dependence on the binding energy $E_{-5}$, which varies from 7.1 to 10.8~GHz across the range shown in Fig.\ \ref{fig:simplified-L0}.

\begin{figure}[tbp]
\centering
\includegraphics[width=0.5\textwidth]{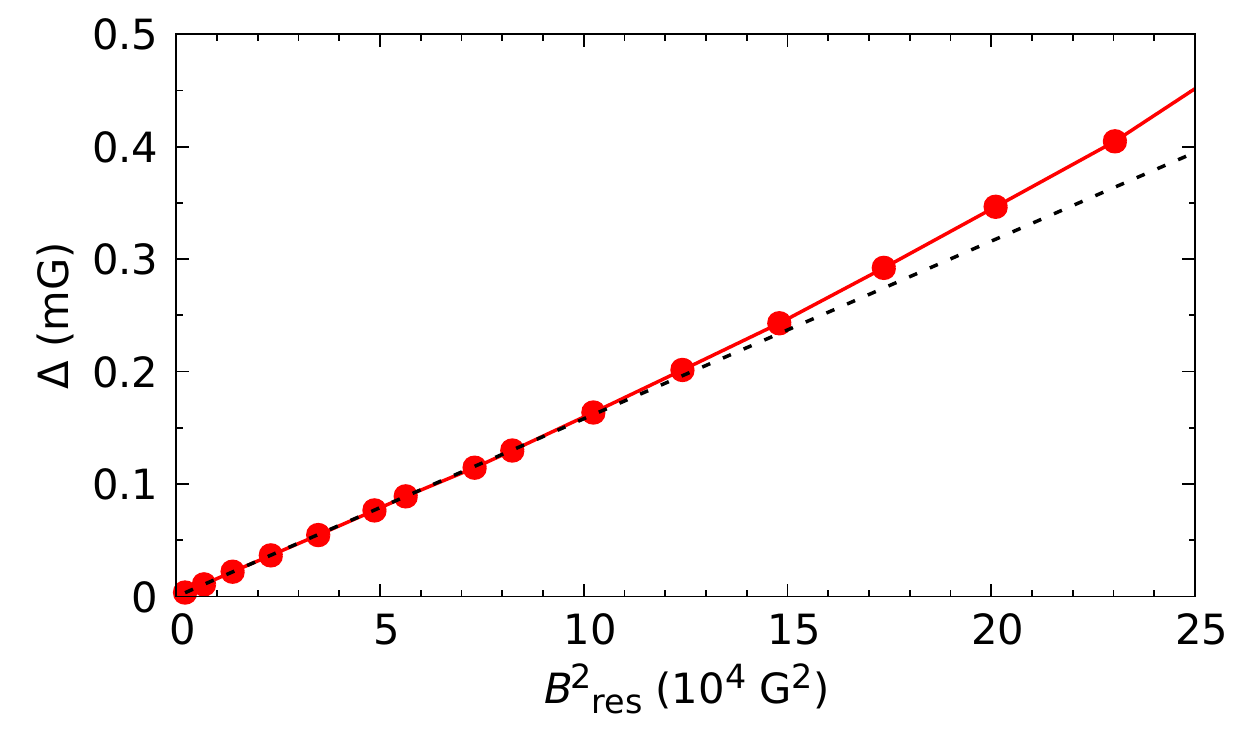}
\caption{Dependence of the width of the resonance in Fig.\ \ref{fig:simplified-L0} on $B_\textrm{res}^2$. The dashed line shows a fit proportional to $B_\textrm{res}^2$ to the widths at low fields.}
\label{fig:L0_vs_B}
\end{figure}

Figure \ref{fig:L0_vs_B} shows the calculated widths as a function of $B_\textrm{res}^2$, with the resonance position tuned using $\lambda_\textrm{scl}$ as in Fig.\ \ref{fig:simplified-L0}. The widths are proportional to $B_\textrm{res}^2$ at low field, but deviate at higher $B_\textrm{res}$, both because of the dependence on $a_\textrm{bg}$ through Eqs.\ \ref{eq:FGRDelta1} and \ref{eq:cqdt} and because the mixing of $f=1$ and $f=2$ is nonlinear in $B$ at higher fields.

\subsection{Resonant states with $L_\textrm{res}=2$}

\begin{figure}[tbp]
\centering
\includegraphics[width=0.5\textwidth]{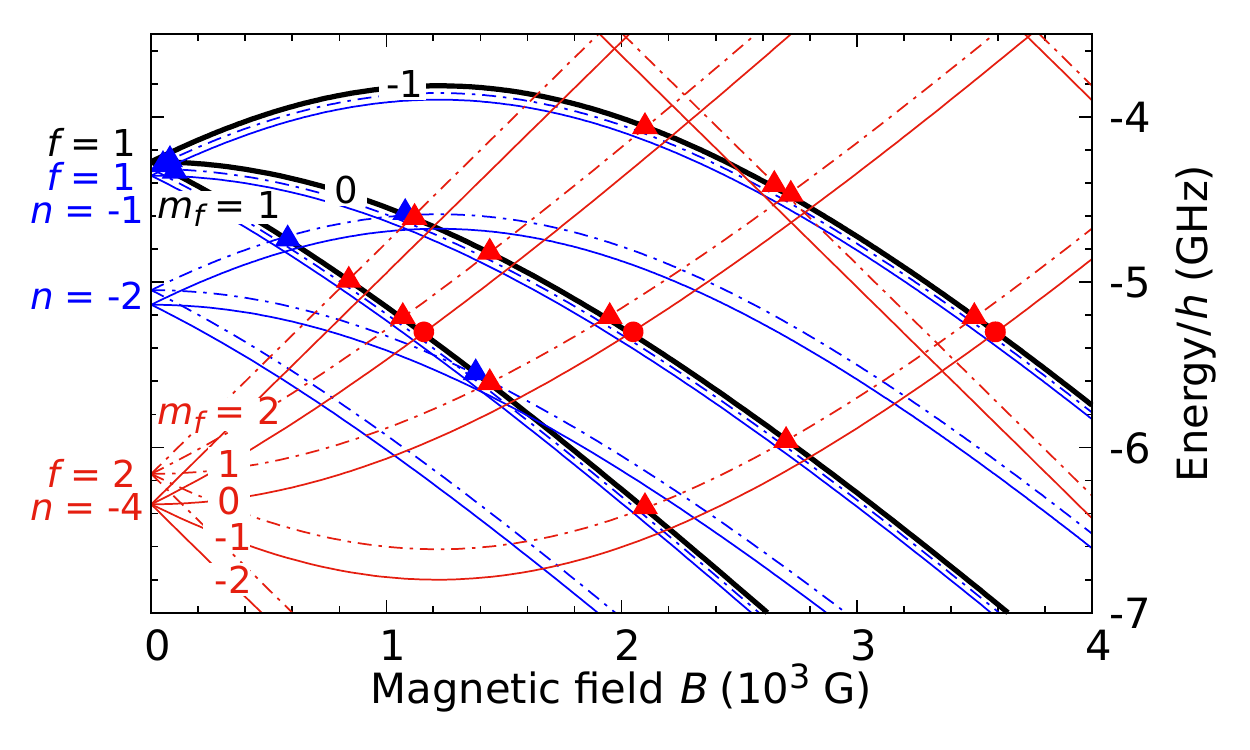}%
\caption{Level crossing diagram for $^{87}$Rb+$^{172}$Yb($^3$P$_0$) with the unscaled interaction potential. The heavy black lines show the hyperfine thresholds for $f = 1$. The quantum numbers $f$ and $n$ are given on the left-hand side for each manifold of molecular levels (thin colored lines, solid for $L = 0$ and dashed for $L = 2$). Crossings that cause Feshbach resonances are indicated with symbols as described in the text.
}%
\label{fig:crossing-L2}
\end{figure}

The resonances due to states with $L_\textrm{res}=2$ (d-wave states) are considerably more complicated. Figure \ref{fig:crossing-L2} shows the crossings between bound states and thresholds for Rb+$^{172}$Yb with the unscaled interaction potential. In this case the states with $L_\textrm{res}=0$ are less deeply bound than those for Rb+$^{174}$Yb in Fig.\ \ref{fig:crossing-L0}, so cause resonances at lower fields, as shown by circles in Fig.\ \ref{fig:crossing-L2} \footnote{The vibrational quantum number is $n=-4$ for Rb+$^{172}$Yb, compared to $n=-5$ for Rb+$^{174}$Yb, because the least-bound state for $^{174}$Yb has become unbound for $^{172}$Yb.}. The states with $L_\textrm{res}=2$ are almost parallel to them as a function of $B$ and slightly higher in energy. However, they produce a more extensive set of resonances in s-wave scattering ($L_\textrm{in}=0$) because conservation of $M_\textrm{tot}$ can be achieved with $|m_{f,\textrm{res}}-m_{f,\textrm{in}}|=0$, 1 or 2, with the change in $m_f$ compensated by a change in $M_L$. The crossings that produce such resonances are indicated with triangles in Fig.\ \ref{fig:crossing-L2}. They can exist even for spin-zero isotopes of atoms in $^3$P$_0$ states; this contrasts with the situation for atoms in $^1$S states, where resonances due to states with $L_\textrm{res}=2$ exist only for isotopes with nuclear spin \cite{Brue:LiYb:2012, Yang:CsYb:2019}.

\begin{figure}[tbp]
\centering
\includegraphics[width=0.5\textwidth]{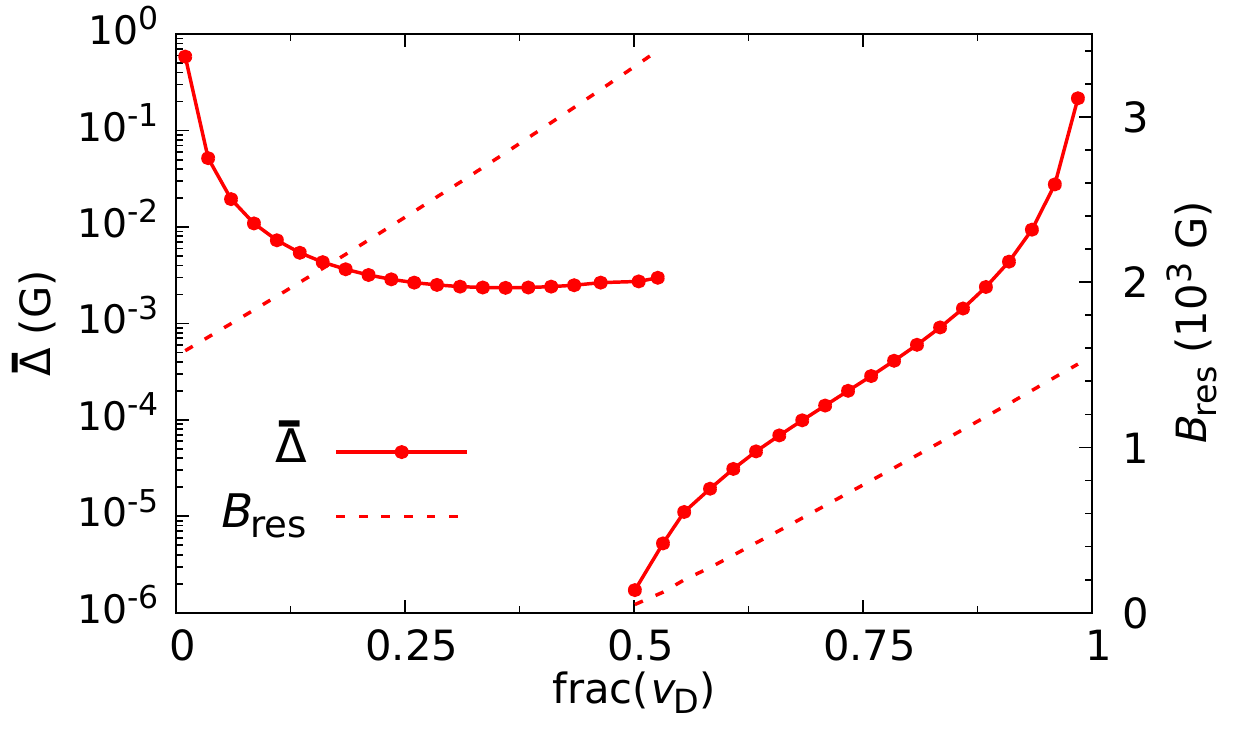}
\caption{Positions $B_\textrm{res}$ (dashed lines) and normalized widths $\bar{\Delta}$ (circles) for resonances with $(f_{\rm res}$, $m_{f,{\rm res}}$, $j_{\rm res}$, $m_{j,{\rm res}}$, $L_{\rm res}$, $M_{L,{\rm res}}) = (2, 2, 0, 0, 2, -1)$.
The quantity frac($v_\textrm{D}$) is related to $a_\textrm{bg}$ through Eq.\ \ref{eq:vD}. The positions and widths are calculated for $^{87}$Rb+$^{174}$Yb using the full Hamiltonian.
}
\label{fig:widths-L2}
\end{figure}

Figure \ref{fig:widths-L2} shows the calculated resonance positions and widths for $(f_{\rm res}, m_{f,{\rm res}}, L_{\rm res}, M_{L,{\rm res}})=(2,2,2,-1)$.
The resonance widths show a similar general dependence on $a_\textrm{bg}$ as those for $L_\textrm{res}=0$ in Fig.\ \ref{fig:simplified-L0}, peaking at large values of $|a_\textrm{bg}|$.

\begin{figure}[tbp]
\centering
\includegraphics[width=0.5\textwidth]{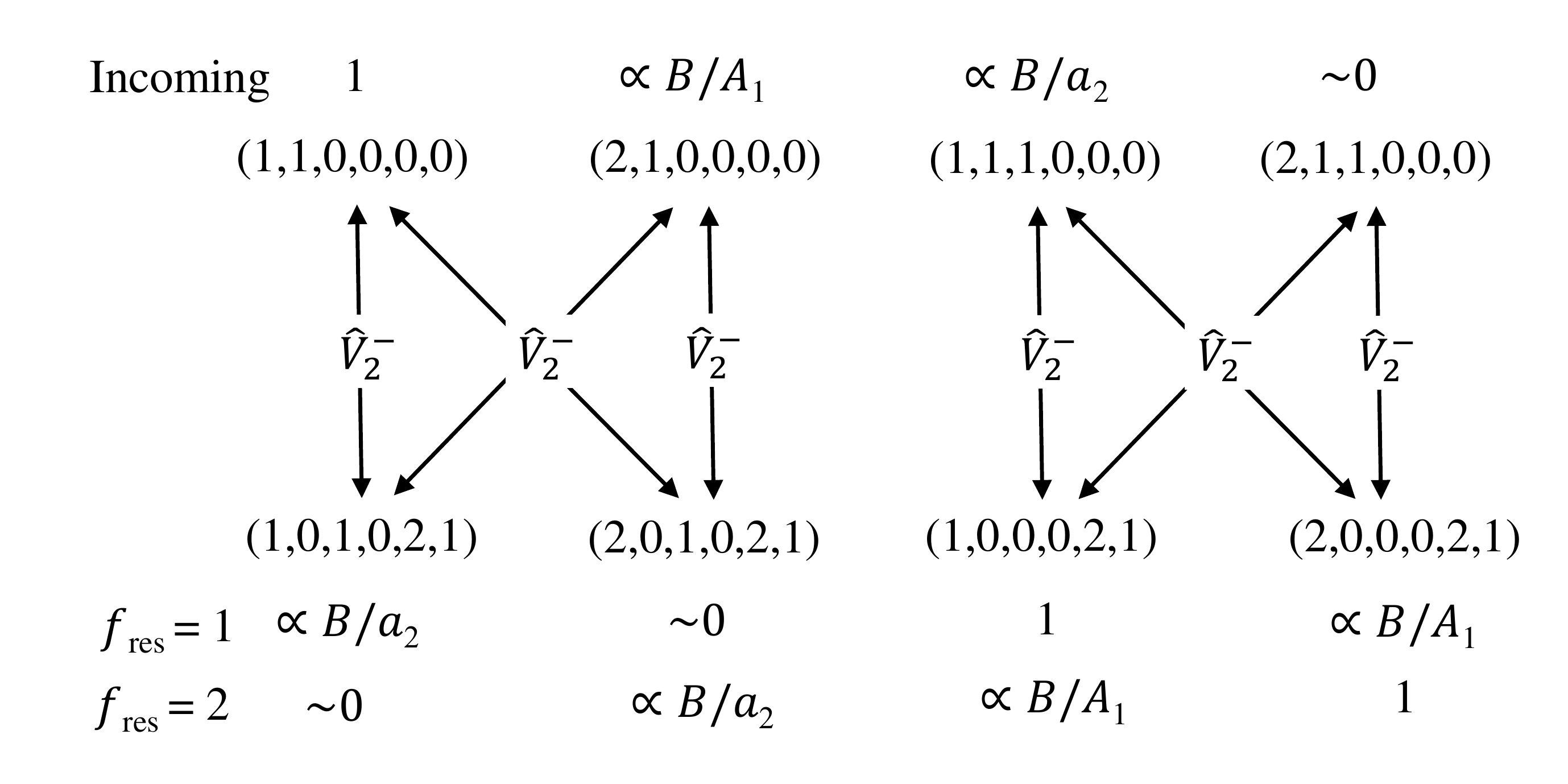}
\caption{Mechanisms of resonances in the incoming channel $(f_\textrm{in},m_{f,\textrm{in}})=(1,1)$ for $^{87}$Rb+Yb($^3$P$_0$), due to bound states with $L_\textrm{res}=2$. Components of the channel functions, designated by $(f,m_f,j,m_j,L,M_L)$, are shown for the incoming and resonant channels, together with the matrix elements between them that drive the resonances. State compositions are shown for resonant states in both the lower hyperfine manifold, with $(f_\textrm{res},m_{f,\textrm{res}})=(1,0)$, and in the upper hyperfine manifold, with $(f_\textrm{res},m_{f,\textrm{res}})=(2,0)$.
}
\label{fig:coupling-L2}
\end{figure}

The lowest-order coupling between the incoming and resonant states is due to the Zeeman term combined with $\hat{V}_2^-$. The compositions of the field-dressed incoming and resonant states, and the couplings between their components due to $\hat{V}_2^-$, are shown in Fig.\ \ref{fig:coupling-L2}. In addition to the direct couplings shown, there are indirect couplings involving the combination of $\hat{V}_2^+$ and $\hat{V}_0^-$ via additional intermediate states. Furthermore, for $L_\textrm{res}=2$ there are $B$-independent terms that can contribute to $\Delta$: these exist even at zero field because the incoming total angular momentum ${\cal F}_\textrm{in}$ is 1 and $j_\textrm{res}=0$, $f_\textrm{res}$ and $L_\textrm{res}=2$ can couple to give several values of ${\cal F}_\textrm{res}$ that include 1. However, their contributions are very small, and $\Delta$ is dominated by terms proportional to $B^2$ at the fields of interest, as for $L_\textrm{res}=0$.

Molecules formed below the Yb($^3$P$_0$) thresholds cannot decay by the mechanisms included in our model, so they have infinite lifetime in the present calculations. There are nevertheless two mechanisms by which they can decay: radiationless decay to form Rb($^2$P)+Yb($^1$S) or Rb($^2$S)+Yb($^1$S), and radiative decay to form Rb($^2$S)+Yb($^1$S). The rate of the radiationless processes is very hard to estimate because they involve spin-orbit coupling matrix elements that cannot be extracted reliably from the electronic structure calculations of ref.\ \cite{Shundalau:2017}. Nevertheless, such processes are expected to be slow because of the large kinetic energy releases involved. In addition, we can estimate the rate of radiative decay. Bound states below the Yb($^3$P$_0$) thresholds have some $^3$P$_1$ character at short range due to the couplings involving $\hat{V}_0^-$ and $\hat{V}_2^-$. The lifetime of the bound state will be approximately  $\tau_{\rm mol}=\tau_{\rm at}/p_1$, where $\tau_{\rm at}$ is the radiative lifetime of Yb($^3$P$_1$) and $p_1$ is the fraction of $^3$P$_1$ character in the bound state. We obtain $p_1=4.79 \times 10^{-3}$ from coupled-channel calculations of the bound-state wavefunction below the resonance at 2070 G for the unscaled potential. Together with $\tau_{\rm at}=850$ ns \cite{Bowers:1996}, this gives $\tau_{\rm mol}=0.18$ ms. This is not a quantitative prediction, but nevertheless indicates that radiative decay will be very slow.

\section{Likelihood of wide resonances at moderate magnetic field}
\label{sec:likely}

\begin{figure}[tbp]
\centering
\includegraphics[width=0.5\textwidth]{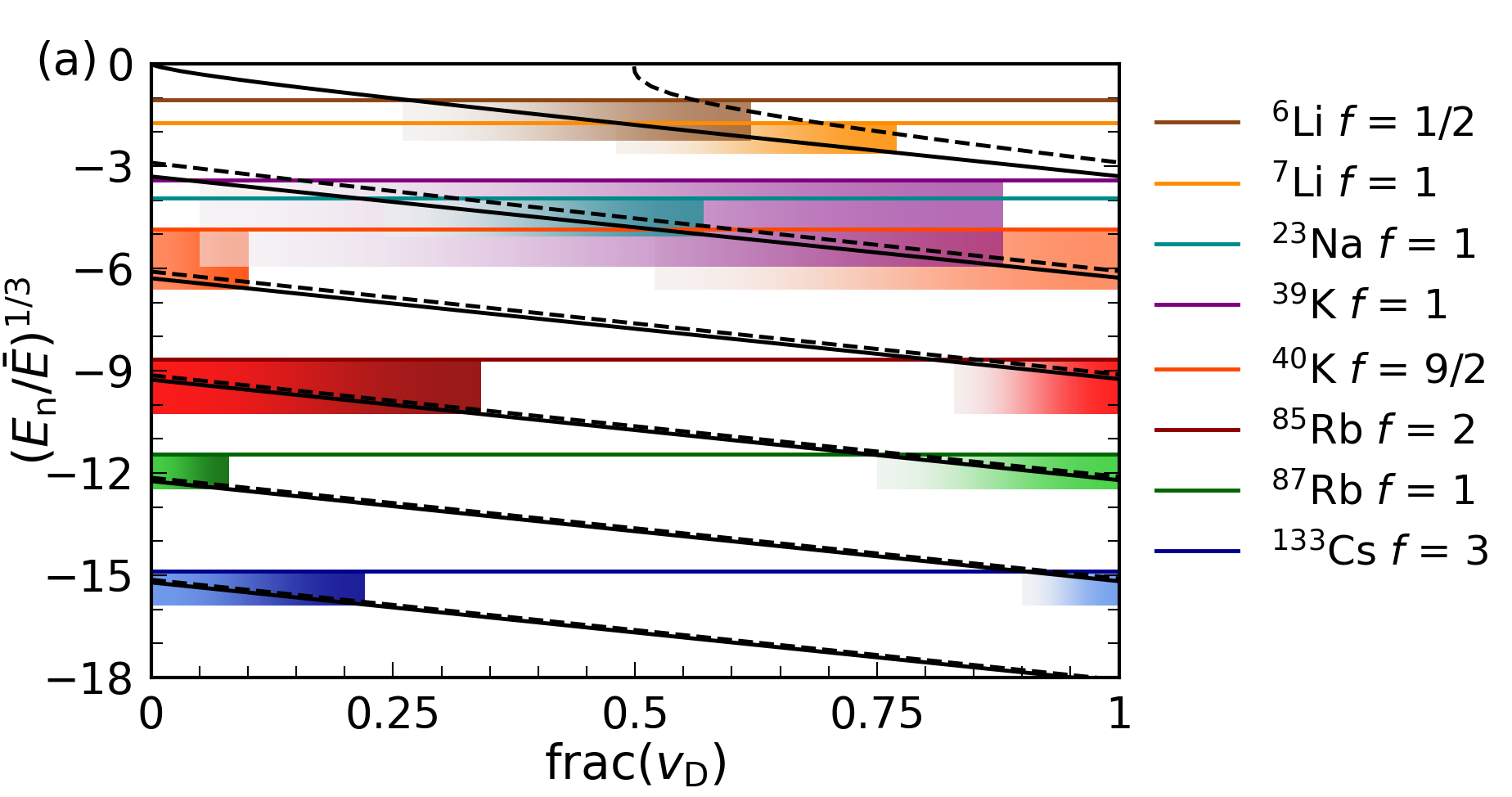}
\includegraphics[width=0.5\textwidth]{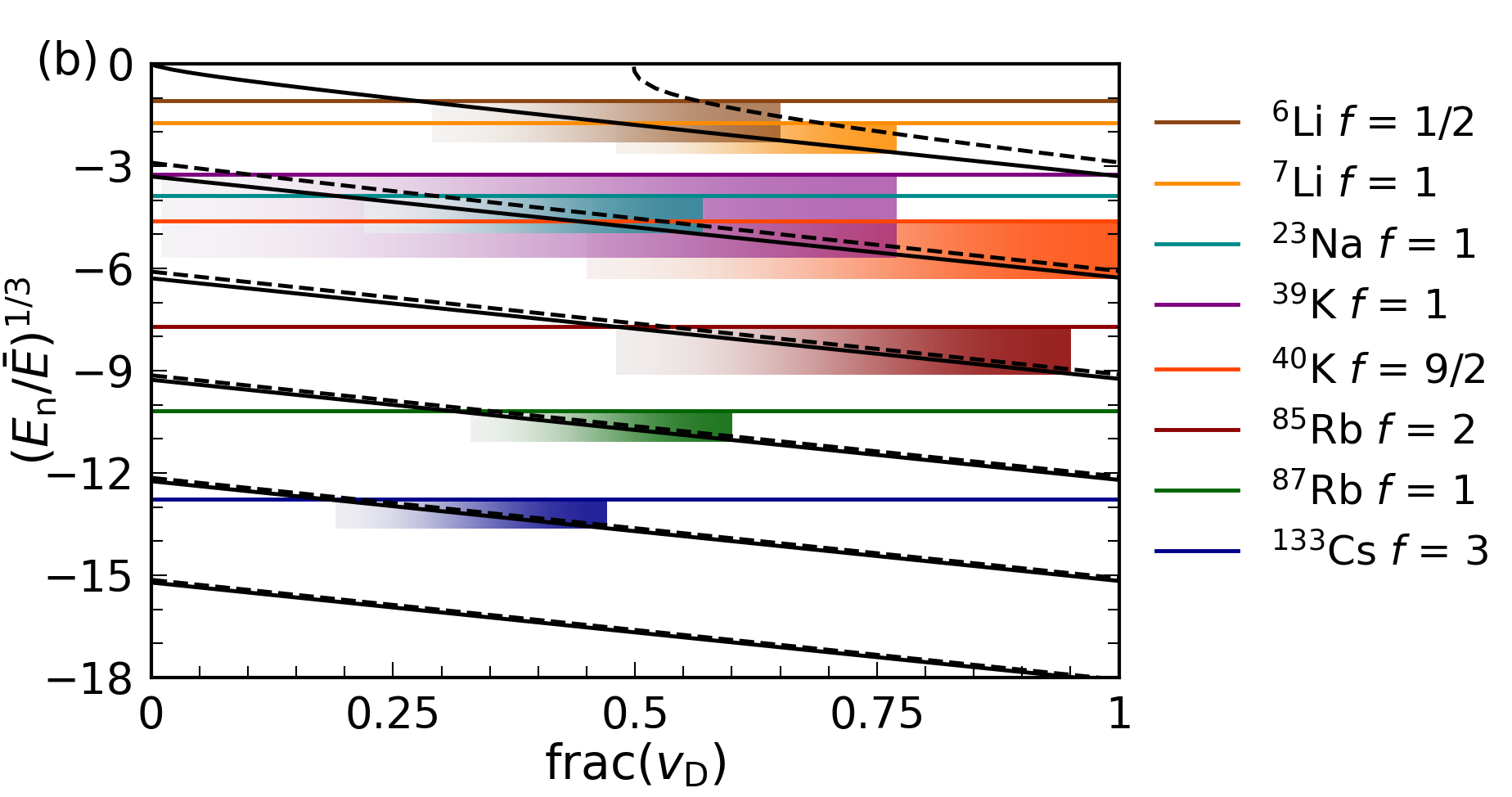}
\caption{(a) Energy levels of $^2$S + Yb($^3$P$_0$) systems, in reduced energy units $\bar{E}$, for different alkali-metal atoms. Colored horizontal lines represent the energy of the lower hyperfine level of each alkali-metal isotope, relative to the upper level. The shaded regions are 2 GHz deep below the lower threshold, and correspond to regions of binding energy where resonances would be expected at fields below about 1000~G; the opacity increases with binding energy below the lower threshold, so that deeper colors indicate higher fields that will increase resonance widths. (b) Similar plot for $^2$S + Sr($^3$P$_0$) systems. In both cases $^{41}$K is omitted because it obscures the band for $^{39}$K, but a version of the Figure including $^{41}$K instead of $^{39}$K is given in Supplemental Material \cite{sup-mat-AlkYb}.}
\label{fig:bound_states}
\end{figure}

A major purpose of the present paper is to consider the relative widths expected for resonances in systems containing different $^2$S and $^3$P atoms. For the $^2$S atoms, we consider the alkali-metal atoms from Li to Cs (9 isotopes in total) and for the $^3$P atoms we consider Yb and Sr, both of which have multiple spin-0 isotopes.

Current cold-atom experiments can comfortably reach magnetic fields around 1000~G, but fields of 2000~G or more are experimentally challenging, particularly when precise field control is needed.
Since the resonance widths contain a factor proportional to $B^2$, it is highly desirable to find systems where resonances appear in a Goldilocks zone, with a field neither too high nor too low.

Similar considerations apply to $a_\textrm{bg}$. The largest widths occur when $|a_\textrm{bg}|$ is large, but large positive $a_\textrm{bg}$ may prevent mixing of condensates, while large negative $a_\textrm{bg}$ may cause condensate collapse. These issues can be circumvented by loading the atoms into an optical lattice or tweezers, but such approaches are challenging in themselves.
As described above, there is also a close connection between $a_\textrm{bg}$ and the binding energies $E_n$ of near-threshold states, which gives rise to a connection between $a_\textrm{bg}$ and the fields at which resonances occur.
We cannot predict $a_\textrm{bg}$ from theory alone; it must be determined from experiments. However, since the two effects described both depend on $a_\textrm{bg}$, we \emph{can} determine whether there are ranges of $a_\textrm{bg}$ for which they are simultaneously favorable for a given system.
Even if the region of the enhancement is quite narrow, $a_\textrm{bg}$ can be tuned to some extent by choosing different isotopes of the $^3$P atom.

For a single-channel system with an asymptotic potential $-C_6R^{-6}$, there is always a single s-wave bound state (with $n=-1$) within $\sim 36\bar{E}$ of threshold, where $\bar{E}=\hbar^2/(2\mu \bar{a}^2)$. This is known as the `top bin', and its width depends on only the reduced mass $\mu$ and the $C_6$ coefficient, but the position of the bound state within the bin depends on $\textrm{frac}(v_\textrm{D})$ (or equivalently on $a_\textrm{bg}$). There are similar bins for $n<-1$, with depths approximately proportional to $(n+\frac{1}{8})^3$. The quantity $(E_n/\bar{E})^{1/3}$ is linear as a function of $\textrm{frac}(v_\textrm{D})$ \cite{LeRoy:1970}, with a small deviation near threshold \cite{Gribakin:1993, Boisseau:1998}. This is shown as a solid line in Fig.\ \ref{fig:bound_states} for a pure $C_6$ potential, with a hard wall to adjust $v_\textrm{D}$. The corresponding quantity for d-wave states is shown as a dashed line. The binding energies are system-independent to the extent that the long-range potential can be represented by $C_6$ alone, with all the scaling encapsulated by different values of $\bar{E}$.

The positions of resonances also involve the hyperfine splitting of the alkali-metal atom. This brings in a different energy scale and introduces system dependence. In Fig.\ \ref{fig:bound_states} the hyperfine splitting is shown by a horizontal colored line for each alkali-metal isotope, scaled by the appropriate $\bar{E}$. The values of $\bar{E}$ are calculated using values of $C_6$ from Tang's combining rule \cite{Tang:1969} together with atomic polarizabilities and homonuclear $C_6$ coefficients \cite{Derevianko:2001, Santra:2004, Dzuba:2010, Guo:SrYb:2010}. The values of $C_6$, $\bar{a}$ and $\bar{E}$ are given in Table \ref{tab:at-prop}. In this representation, the bound states are those from channels of the upper hyperfine state and the horizontal lines are the zero-field energies of the lower hyperfine state. Low-field resonances for a specific system thus occur for values of $\textrm{frac}(v_\textrm{D})$ where there is a bound state just below its corresponding horizontal line. However, distances below the horizontal lines scale differently for different systems, due to both the $E_n^{1/3}$ scaling of the axis and different values of $\bar{E}$. We therefore also show shaded regions covering 2 GHz below each line, placed to show where there is a state in this window. For $^{87}$Rb, a state at the bottom of this region will cause a resonance with $m_{f,\textrm{res}}=m_{f,\textrm{in}}=1$ at about 1120~G; the corresponding field for other alkali-metal isotopes is included in Table \ref{tab:at-prop}.
The d-wave states (dashed lines in Fig.\ \ref{fig:bound_states}) are close to the corresponding s-wave states on this scale, so will cause resonances for similar ranges of $a_\textrm{bg}$. They will cause additional resonances with $m_{f,\textrm{res}}>m_{f,\textrm{in}}$ at somewhat lower magnetic fields, which may be more accessible.

\begin{table*}[tbp]
\caption{Dispersion coefficients $C_6$, mean scattering lengths $\bar{a}$ and corresponding energies $\bar{E}$ for combinations of alkali-metal atoms with Yb and Sr in $^3$P$_0$ states. Values are given for $^{172}$Yb and $^{86}$Sr, but are very similar for other isotopes. Also given is the magnetic field $B_\textrm{2\,GHz}$ at which the two hyperfine states of the alkali-metal atom with $m_f=i-\frac{1}{2}$ (or $-i+\frac{1}{2}$ for $^{40}$K) are separated by 2~GHz more than at zero field. \label{tab:at-prop}} \centering
\begin{ruledtabular}
\begin{tabular}{cccccccc}
%\hline\hline
& \multicolumn{3}{c}{$^{172}$Yb} & \multicolumn{3}{c}{$^{86}$Sr} & \\
\cline{2-4}\cline{5-7}
%\hline
& $C_6$ ($E_\textrm{h}a_0^6$)	& $\bar{a}$ ($a_0$)	& $\bar{E}/h$ (MHz)
& $C_6$ ($E_\textrm{h}a_0^6$)	& $\bar{a}$ ($a_0$)	& $\bar{E}/h$ (MHz) & $B_\textrm{2\,GHz}$ (G) \\
\hline
$^6$Li     & 2313 & 40.0 & 194  & 2620 & 40.9 & 192  & 760 \\
$^7$Li     & 2313 & 41.5 & 155  & 2620 & 42.4 & 155  & 830 \\
$^{23}$Na  & 2427 & 55.3 & 29.1 & 2724 & 55.4 & 32.4 & 910 \\
$^{39}$K   & 3888 & 69.6 & 11.7 & 4442 & 69.0 & 14.1 & 780 \\
$^{40}$K   & 3888 & 70.0 & 11.4 & 4442 & 69.3 & 13.8 & 780 \\
$^{41}$K   & 3888 & 70.3 & 11.0 & 4442 & 69.6 & 13.4 & 760 \\
$^{85}$Rb  & 4266 & 82.4 & 4.67 & 4874 & 79.3 & 6.71 & 880 \\
$^{87}$Rb  & 4266 & 82.7 & 4.57 & 4874 & 79.6 & 6.60 & 1120\\
$^{133}$Cs & 5158 & 92.6 & 2.81 & 5928 & 87.6 & 4.51 & 890 \\
%\hline\hline
\end{tabular}
\end{ruledtabular}
\end{table*}

The most favorable resonances will occur when both $|a_\textrm{bg}|$ and $B_\textrm{res}$ are large but not too large. On Fig.\ \ref{fig:bound_states}, the former occurs near either end of the horizontal axis, while the latter corresponds to regions near the right-hand (darker) end of the appropriate shaded bar. For the Yb systems (Fig.\ \ref{fig:bound_states}(a)), it can be seen that there are regions that match both these criteria well for $^{40}$K, $^{87}$Rb and $^{133}$Cs; for the Sr systems (Fig.\ \ref{fig:bound_states}(b)), $^{40}$K and $^{85}$Rb look the most promising, although the right-hand edges of the shaded regions shown in Fig.\ \ref{fig:bound_states}(b) correspond to large negative scattering lengths, which may pose experimental challenges.

Higher-order terms in the potentials will also affect these results, particularly for the deeper states. Inclusion of the next two dispersion terms $-C_8R^{-8}-C_{10}R^{-10}$ with plausible strengths shifts the crossing point for $^{133}$Cs+Yb to lower $v_\textrm{D}$ by about 0.1 and proportionately less for lighter atoms.

\begin{table}[tbp]
\caption{Variation $\Delta\mu$ of reduced mass across the range of available isotopes for Yb and Sr, together with the resulting change in $\Delta v_\textrm{D}/v_\textrm{D}$, for different alkali-metal atoms.\label{tab:scaling}} \centering
\begin{ruledtabular}
\begin{tabular}{ccccc}
%\hline\hline
		& \multicolumn{2}{c}{$^{168}$Yb to $^{176}$Yb}	 & \multicolumn{2}{c}{$^{84}$Sr to $^{88}$Sr} \\
		\cline{2-3}\cline{4-5}
	& $\Delta \mu$ (u)	& $\Delta v_\textrm{D}/v_\textrm{D}$ (\%)	& $\Delta \mu$ (u)	& $\Delta v_\textrm{D}/v_\textrm{D}$ (\%) \\ \hline
 $^6$Li  & 0.01 & 0.08 & 0.02 & 0.15 \\
 $^7$Li  & 0.01 & 0.09 & 0.02 & 0.18 \\
 $^{23}$Na  & 0.11 & 0.28 & 0.18 & 0.49 \\
 $^{39}$K  & 0.27 & 0.43 & 0.39 & 0.73 \\
 $^{40}$K  & 0.28 & 0.44 & 0.40 & 0.74 \\
 $^{41}$K  & 0.30 & 0.45 & 0.42 & 0.75 \\
 $^{85}$Rb  & 0.88 & 0.77 & 0.99 & 1.16 \\
 $^{87}$Rb  & 0.90 & 0.79 & 1.01 & 1.18 \\
 $^{133}$Cs  & 1.52 & 1.02 & 1.47 & 1.42\\
% \hline\hline
\end{tabular}
\end{ruledtabular}
\end{table}

We now consider how much $v_\textrm{D}$, or equivalently $a_\textrm{bg}$, can be tuned for different systems by changing the isotope of the $^3$P atom. Yb has stable isotopes from $^{168}$Yb to $^{176}$Yb and Sr from $^{84}$Sr to $^{88}$Sr. $v_\textrm{D}$ is approximately proportional to $\mu^{1/2}$. Table \ref{tab:scaling} shows the tuning of reduced mass, and the percentage tuning of $v_\textrm{D}$ possible for various systems. This demonstrates the degree to which a large mass ratio between $^2$S and $^3$P atoms inhibits the tuning possible.
The present potential for Rb+Yb($^3$P$_0$) supports 136 bound states, so varying the Yb isotope allows tuning of $v_\textrm{D}$ by $\sim 1.1$, covering more than a full cycle of scattering length. Interaction potentials and numbers of bound states are unavailable for most other systems. However, by analogy with the potentials for alkali-metal atoms with ground-state Yb \cite{Brue:AlkYb:2013}, we expect that the number of bound states, and so the tuning, will be significantly smaller for the lighter atoms. This again emphasises the benefit of working with the heavier $^2$S atoms when substantial tuning of the scattering length with isotope is desired.

\section{Conclusions}
\label{sec:conclude}

We have investigated the mechanisms of magnetically tunable Feshbach resonances between alkali-metal atoms and atoms such as Yb and Sr in $^3$P$_0$ states. We have found that resonances due to s-wave bound states are driven by a combination of the Zeeman effect and a component of the potential $\hat{V}_0^-$ that characterizes the isotropic part of the difference between the singlet and triplet electronic states. Because of this, the widths of the resonances are to a good approximation quadratic in magnetic field $B$. We have also investigated the dependence of the widths on the background scattering length $a_\textrm{bg}$; there is a very strong dependence, peaking when the magnitude of $a_\textrm{bg}$ is large.

Resonances due to d-wave bound states can also occur, and there are more of them than those due to s-wave states. Their mechanisms are more complicated, but they are mostly driven by an analogous mechanism involving the anisotropic part of the difference potential, $\hat{V}_2^-$. Their widths have generally similar dependence on magnetic field and $a_\textrm{bg}$.

The mechanisms for both s-wave and d-wave states involve mixing between $^3$P$_0$ and $^3$P$_1$ states due to the difference potential. The resonances are therefore generally expected to be broader for Sr than for Yb, because of its smaller spin-orbit splitting.

The resonances are generally narrow except when they occur for an isotopic combination with a large magnitude of $a_\textrm{bg}$. Fortunately, both Sr and Yb have several isotopes, and the differing reduced masses offer some discrete tuning of $a_\textrm{bg}$.  We have quantified the extent of this tuning: for heavier alkali-metal atoms (Rb and Cs), $a_\textrm{bg}$ can be tuned over most or all of a complete cycle from $+\infty$ to $-\infty$.

A further consideration involves the magnetic field. The broadest resonances occur at large $B$. However, precise magnetic field control is experimentally challenging at fields much above 1000~G. We have developed a quantitative picture, based on the pattern of near-threshold levels, to identify combinations of alkali-metal and $^3$P atoms that can have large background scattering lengths at the same time as resonant fields at the upper end of the accessible range. Particularly promising systems include $^{87}$Rb+Yb, Cs+Yb and $^{85}$Rb+Sr.

Analogous resonances will exist in systems made up of an alkali-metal atom and an atom such as Cd or Hg in a $^3$P state.

The work described here paves the way for a new approach to making polar molecules in $^2\Sigma$ states by magnetoassociation followed by laser transfer to the ground state. Such molecules have important potential applications in a variety of fields, ranging from quantum simulation to the testing of fundamental symmetries of nature.

\section*{Data availability statement}

The data presented in this work are available from Durham University
\cite{DOI_data-AlkYb}.

\begin{acknowledgments}
This work was supported by the U.K. Engineering and Physical Sciences Research Council (EPSRC) Grant EP/P01058X/1.
\end{acknowledgments}

\bibliographystyle{long_bib}
\bibliography{../all,AlkYb3PData}
\end{document}